\documentstyle[preprint,aps]{revtex}

\begin{document}
\draft
\author{Bao-Sen Shi\thanks{%
E-mail: drshi@ustc.edu.cn}, Yun-Kun Jiang and Guang-Can Guo\thanks{%
E-mail: gcguo@ustc.edu.cn}}
\address{Lab. of Quantum Communication and Quantum Computation \\
Department of Physics \\
University of Science and Technology of China\\
Hefei, 230026 P. R. China}
\title{Manipulating the frequency entangled states by acoustic-optical-modulator}
\maketitle

\begin{abstract}
In this paper, we describe how to realize conditional frequency entanglement
swapping and to produce probabilisticly a three-photon frequency entangled
state from two pairs of frequency entangled states by using an
Acoustic-Optical-Modulator. Both schemes are very simple and may be
implementable in practice.
\end{abstract}


\section{Introduction}

Ever since the seminal work of Einstein, Podolsky, and Rosen [1] there has
been a quest for generating entanglement between quantum particles. The
resource of entanglement [2] has many useful applications in quantum
information processing, such as secret key distribution [3], quantum
teleportation [4] , dense coding [5], and so on. Two polarization entangled
photons, which are produced by spontaneous parametric down-conversion (SPDC)
with type II phase matching in nonlinear crystal [6] have been used to
realize both dense coding [7] and quantum teleportation [8]. Entanglement
swapping [9], which enables one to entangle two quantum systems that have
never interacted directly with each other, has been demonstrated
experimentally [10]. Maximally entangled state of three or more particles,
so called Greenberger-Horne-Zeilinger (GHZ) state [11], has been fascinating
quantum systems to reveal the nonlocality of the quantum world. There are
some methods to produce the maximally entangled multiply particles
states[12, 13]. The proposals for three particles entangled states have been
made for experiments with photons [13] and atoms [14]. Three nuclear spins
within a single molecule have been prepared such that they locally exhibit
three-particle correlations [15]. Recently, Bouwmeester {\it et al} [16]
have observed successfully the entanglement of three-photon GHZ state. The
main idea of this scheme is to transform two pairs of polarization entangled
photons into three polarization entangled photons and a fourth independent
photon. In all above proposals, only polarization or momentum entanglement
of multiply particles is considered. The frequencies or energies of these
particles in an entangled state are assumed to be equal. Molotkov and Nazin
[17] considered the case of the frequency entangled state and presented a
simple nonlinear crystal based optical scheme for experimental realization
of the frequency entanglement swapping between the photons belonging to
independent biphotons. But this scheme is unpractical because of the lower
nonlinear susceptibility of the nonlinear crystal. In this paper, the
frequency entangled state is considered. We find that the frequency
entanglement swapping can be realized only by Acoustic-Optical-Modulator
(AOM). In contrast to the scheme of Ref. [10], no Bell state measurement is
needed. Furthermore, we can produce a three-photon frequency entangled state
from two pairs of frequency entangled states by AOM. These schemes are
simply and may be implementable in practice.One disadvantage of our scheme
is both schemes are probabilisticly realizable.

The paper is outlined as follows. In Sec. II, we give the frequency
transformation done by AOM; In Sec. III, we present a proposal for frequency
entanglement swapping; In Sec. IV, we show how to produce a three-photon
frequency entangled state from two pairs of frequency entangled states;
Finally, we give a brief conclusion in Sec. V.

\section{frequency transformation done by AOM}

In this paper, all proposals are realized by AOM, so before proceeding, we
will give the frequency transformation done by AOM. Suppose there is an AOM
(for example, Acousto-Optic Brag cell), which is driven at radio frequency
(rf) $\delta $. If a monochromatic beam of frequency $\omega $ is introduced
into the AOM, then it will be separated into two beams, one is a transmitted
beam and the other is a diffracted beam. The frequency of the transmitted
beam holds unchanged and the frequency of the diffracted wave shifts. The
shift may be $\delta $ or -$\delta $, depending on the directions of the
incident beam and the sound wave in AOM. For example, in Fig. 1, if the
incident beam transmits along the path 1 (1$^{^{\prime }}$), and the
direction of the sound wave in AOM is shown by arrow $\rightarrow $, the
diffracted beam $d(t)$ will experience a frequency shift of $\delta $ (-$%
\delta $).[18]

The amplitudes of the transmitted and diffracted waves can be adjusted by
the amplitude of the sound wave. If only one photon of frequency $\omega $
enters this AOM along path 1, and we adjust the amplitude of the sound $%
\delta $ so that the amplitudes of the transmitted and diffracted waves are
equal, then the transformation done by AOM can be written as the following:

\begin{equation}
\left| \omega \right\rangle _1\stackrel{AOM}{\rightarrow }\frac 1{\sqrt{2}}%
[\left| \omega \right\rangle _t+\left| \omega +\delta \right\rangle _d]. 
\eqnum{1}
\end{equation}
If a photon of frequency $\omega +\delta $ enters AOM along path 1$%
^{^{\prime }}$, the transformation done by AOM is

\begin{equation}
\left| \omega \right\rangle _{1^{^{\prime }}}\stackrel{AOM}{\rightarrow }%
\frac 1{\sqrt{2}}[\left| \omega \right\rangle _t+\left| \omega +\delta
\right\rangle _d].  \eqnum{2}
\end{equation}
With the desmontration above in mind, we proceed to the following.

\section{Photon frequency entanglement swapping}

Suppose we are given two biphotons frequency entangled states $\left| \Phi
\right\rangle $ and $\left| \Psi \right\rangle $:

\begin{equation}
\left| \Phi \right\rangle =\frac 1{\sqrt{2}}[\left| \omega \right\rangle
_1\left| \omega +\delta \right\rangle _2+\left| \omega +\delta \right\rangle
_{1^{^{\prime }}}\left| \omega \right\rangle _{2^{^{\prime }}}]  \eqnum{3}
\end{equation}

and

\begin{equation}
\left| \Psi \right\rangle =\frac 1{\sqrt{2}}[\left| \omega \right\rangle
_3\left| \omega +\delta \right\rangle _4+\left| \omega +\delta \right\rangle
_{3^{^{\prime }}}\left| \omega \right\rangle _{4^{^{\prime }}}],  \eqnum{4}
\end{equation}

Where, the subscripts represent beams taken by photons. In an experiment,
these states can be easily obtained for example by SPDC in type I with
noncolinear and nondegenerate phase matching. Obviously, photons in state $%
\left| \Phi \right\rangle $ are independent of photons in state $\left| \Psi
\right\rangle $. In order to make one of photon in $\left| \Phi
\right\rangle $ state entangle with one of photon in state $\left| \Psi
\right\rangle $, we consider the arrangement of Fig.2, in which, two AOMs
are needed. Beams 2 and 3 enter AOM1 driven at a radio frequency $\delta $,
and  beams 2$^{^{\prime }}$ and 3$^{^{\prime }}$ enter AOM2, which is also
driven at the radio frequency $\delta $. We arrange the diffracted beam of
the frequency $\omega $ along the transmitted beam of the frequency $\omega
+\delta $ in each AOM. According to Sec. II, if the frequencies of the beams
2, 2$^{^{\prime }},$ 3 and 3$^{^{\prime }}$ are $\left| \omega +\delta
\right\rangle _2$, $\left| \omega \right\rangle $ $_{2^{^{\prime }}},$ $%
\left| \omega \right\rangle _3$ and $\left| \omega +\delta \right\rangle
_{3^{^{\prime }}}$ respectively, then the following transformations are
obtained:

\begin{equation}
\left| \omega +\delta \right\rangle _2\stackrel{AOM1}{\rightarrow }\frac 1{%
\sqrt{2}}[\left| \omega +\delta \right\rangle _{T_1}+\left| \omega
\right\rangle _{T_1^{^{\prime }}}],  \eqnum{5}
\end{equation}

\begin{equation}
\left| \omega \right\rangle _{2^{^{\prime }}}\stackrel{AOM2}{\rightarrow }%
\frac 1{\sqrt{2}}[\left| \omega \right\rangle _{T_2}+\left| \omega +\delta
\right\rangle _{T_2^{^{\prime }}}],  \eqnum{6}
\end{equation}

\begin{equation}
\left| \omega \right\rangle _3\stackrel{AOM1}{\rightarrow }\frac 1{\sqrt{2}}%
[\left| \omega +\delta \right\rangle _{T_1}+\left| \omega \right\rangle
_{T_1^{^{\prime }}}],  \eqnum{7}
\end{equation}

\begin{equation}
\left| \omega +\delta \right\rangle _{3^{^{\prime }}}\stackrel{AOM2}{%
\rightarrow }\frac 1{\sqrt{2}}[\left| \omega \right\rangle _{T_2}+\left|
\omega +\delta \right\rangle _{T_2^{^{\prime }}}].  \eqnum{8}
\end{equation}

where, $T_1,T_1^{^{\prime }}$ are the directions of transmitted (diffracted)
and diffracted (transmitted) beams of the incident wave $\left| \omega
+\delta \right\rangle _2$ ($\left| \omega \right\rangle _3$) through AOM 1 , 
$T_2,T_2^{^{\prime }}$ are the directions of transmitted (diffracted) and
diffracted (transmitted) beams of the incident wave $\left| \omega
\right\rangle _{2^{^{\prime }}}$ ($\left| \omega +\delta \right\rangle
_{3^{^{\prime }}}$) through AOM 2.

If the initial state of four photons to be $\left| \Phi \right\rangle
\otimes \left| \Psi \right\rangle $, it will transform into:

\begin{eqnarray}
\left| \Phi \right\rangle \otimes \left| \Psi \right\rangle  &=&\frac 12%
\{\left| \omega \right\rangle _1\left| \omega +\delta \right\rangle _4\frac 1%
{\sqrt{2}}[\left| \omega +\delta \right\rangle _{T_1}+\left| \omega
\right\rangle _{T_1^{^{\prime }}}]\otimes \frac 1{\sqrt{2}}[\left| \omega
+\delta \right\rangle _{T_1}+\left| \omega \right\rangle _{T_1^{^{\prime
}}}]+  \nonumber \\
&&\ \left| \omega \right\rangle _1\left| \omega \right\rangle _{4^{^{\prime
}}}\frac 1{\sqrt{2}}[\left| \omega +\delta \right\rangle _{T_1}+\left|
\omega \right\rangle _{T_1^{^{\prime }}}]\otimes \frac 1{\sqrt{2}}[\left|
\omega \right\rangle _{T_2}+\left| \omega +\delta \right\rangle
_{T_2^{^{\prime }}}]+  \eqnum{9} \\
&&\ \left| \omega +\delta \right\rangle _{1^{^{\prime }}}\left| \omega
+\delta \right\rangle _4\frac 1{\sqrt{2}}[\left| \omega +\delta
\right\rangle _{T_1}+\left| \omega \right\rangle _{T_1^{^{\prime }}}]\otimes 
\frac 1{\sqrt{2}}[\left| \omega \right\rangle _{T_2}+\left| \omega +\delta
\right\rangle _{T_2^{^{\prime }}}]+  \nonumber \\
&&\ \left| \omega +\delta \right\rangle _{1^{^{\prime }}}\left| \omega
\right\rangle _{4^{^{\prime }}}\frac 1{\sqrt{2}}[\left| \omega \right\rangle
_{T_2}+\left| \omega +\delta \right\rangle _{T_2^{^{\prime }}}]\otimes \frac 
1{\sqrt{2}}[\left| \omega \right\rangle _{T_2}+\left| \omega +\delta
\right\rangle _{T_2^{^{\prime }}}]\}.  \nonumber
\end{eqnarray}

The first and the last terms of the right of Eq. (9) mean that there are two
photons which transmit through the AOM1 or AOM2, we discard these two cases.
The other two terms mean that there is one photon which enters the AOM1 and
another photon enters the AOM2 respectively, we only discuss these two
terms. Obviously, these two terms can be rewritten as the following:

\begin{equation}
\frac 12[\left| \omega \right\rangle _1\left| \omega \right\rangle
_{4^{^{\prime }}}+\left| \omega +\delta \right\rangle _{1^{^{\prime
}}}\left| \omega +\delta \right\rangle _4]\frac 1{\sqrt{2}}[\left| \omega
+\delta \right\rangle _{T_1}+\left| \omega \right\rangle _{T_1^{^{\prime
}}}]\otimes \frac 1{\sqrt{2}}[\left| \omega \right\rangle _{T_2}+\left|
\omega +\delta \right\rangle _{T_2^{^{\prime }}}].  \eqnum{10}
\end{equation}

For example, if one photon enters AOM1 and at the same time another photon
enters AOM2, photon 1 and photon 4 will be projected to a maximally
frequency entangled state. The effect of AOMs in our scheme is that erasing
all information about frequency and paths taken by photons. The probability
of success is only 50\%. This can be regarded as the probabilistic
entanglement swapping. One advantage of this scheme is no Bell state
measurement is needed, which may make this scheme implementable easily in
practice. One disadvantage of our scheme is photons 2 and 3 are completely
disentangled pair if successful entanglement swapping occurs, which is part
of entanglement loses. Fourthermore, this scheme can be used to realize
photon frequency entanglement purifing  in Ref. [19].

\section{Producing a three-photon frequency entangled state from two pairs
of frequency entangled states}

In order to produce a three-photon frequency entangled state , we consider
the arrangement of Fig.3, in which, only one AOM driven at rf $\delta $ is
needed. Suppose we are also given two biphotons frequency entangled states $%
\left| \Phi \right\rangle $ and $\left| \Psi \right\rangle $ shown in Eqs.
(1) and (2). The beams 2 and 3 enter the AOM . If the frequencies of beams 2
and 3 are $\omega +\delta $ and $\omega $ respectively, according to Sec.
II, the following transformations can be obtained 
\begin{equation}
\left| \omega +\delta \right\rangle \rightarrow \frac 1{\sqrt{2}}[\left|
\omega \right\rangle _T+\left| \omega +\delta \right\rangle _{T^{^{\prime
}}}],  \eqnum{11}
\end{equation}

\begin{equation}
\left| \omega \right\rangle \rightarrow \frac 1{\sqrt{2}}[\left| \omega
\right\rangle _T+\left| \omega +\delta \right\rangle _{T^{^{\prime }}}], 
\eqnum{12}
\end{equation}
where, $T$ and $T^{^{\prime }}$ are the directions of transmitted
(diffracted) and diffracted (transmitted) beams of the incident beam 3 (beam
2). This kind of change can be obtained by the following: arranging the
diffracted beam of the frequency $\omega $ along the transmitted beam of the
frequency $\omega +\delta $.

Obviously, if one of these two detectors D$_T$ and D$_{T^{^{\prime }}}$
detects a photon, we can not get any information from which source this
photon comes by frequency, i.e., information about the source of this photon
by frequency is erased. This make the state of other three photons collapse
into a superposition state. If all information about the source of this
photon is erased, a three-photon frequency entangled state will be obtained.

Now, we discuss this scheme in detail. The product of state $\left| \Phi
\right\rangle \otimes \left| \Psi \right\rangle $ is 

\begin{eqnarray}
&&\ \frac 12\{\left| \omega \right\rangle _1\left| \omega +\delta
\right\rangle _2\left| \omega \right\rangle _3\left| \omega +\delta
\right\rangle _4+\left| \omega \right\rangle _1\left| \omega +\delta
\right\rangle _2\left| \omega +\delta \right\rangle _{3^{^{\prime }}}\left|
\omega \right\rangle _{4^{^{\prime }}}+  \nonumber \\
&&\ \left| \omega +\delta \right\rangle _{1^{^{\prime }}}\left| \omega
\right\rangle _{2^{^{\prime }}}\left| \omega \right\rangle _3\left| \omega
+\delta \right\rangle _4+\left| \omega +\delta \right\rangle _{1^{^{\prime
}}}\left| \omega \right\rangle _{2^{^{\prime }}}\left| \omega +\delta
\right\rangle _{3^{^{\prime }}}\left| \omega \right\rangle _{4^{^{\prime
}}}\}.  \eqnum{13}
\end{eqnarray}
The first term of Eq. (13) means that there is a photon in beam 2 and beam 3
respectively, the fourth term means that there is no photon in both beams.
We discard these two cases (which can be distinguished from the other cases
by the following: observe whether both detectors fire, or both detectors are
dark, or not) and only discuss the other two terms. 
\begin{equation}
\frac 12\{\left| \omega \right\rangle _1\left| \omega +\delta \right\rangle
_2\left| \omega +\delta \right\rangle _{3^{^{\prime }}}\left| \omega
\right\rangle _{4^{^{\prime }}}+\left| \omega +\delta \right\rangle
_{1^{^{\prime }}}\left| \omega \right\rangle _{2^{^{\prime }}}\left| \omega
\right\rangle _3\left| \omega +\delta \right\rangle _4\}.  \eqnum{14}
\end{equation}
By AOM driven at $\delta $, Eq. (14) changes into:

\begin{eqnarray}
&&\frac 1{2\sqrt{2}}\{\left| \omega \right\rangle _1\left| \omega
\right\rangle _T\left| \omega +\delta \right\rangle _{3^{^{\prime }}}\left|
\omega \right\rangle _{4^{^{\prime }}}+\left| \omega +\delta \right\rangle
_{1^{^{\prime }}}\left| \omega \right\rangle _{2^{^{\prime }}}\left| \omega
\right\rangle _T\left| \omega +\delta \right\rangle _4+  \nonumber \\
&&\left| \omega \right\rangle _1\left| \omega +\delta \right\rangle
_{T^{^{\prime }}}\left| \omega +\delta \right\rangle _{3^{^{\prime }}}\left|
\omega \right\rangle _{4^{^{\prime }}}+\left| \omega +\delta \right\rangle
_{1^{^{\prime }}}\left| \omega \right\rangle _{2^{^{\prime }}}\left| \omega
+\delta \right\rangle _{T^{^{\prime }}}\left| \omega +\delta \right\rangle
_4\}.  \eqnum{15}
\end{eqnarray}
Equation (15) can be rewritten as:

\begin{eqnarray}
&&\ \frac 1{2\sqrt{2}}\{[\left| \omega \right\rangle _1\left| \omega +\delta
\right\rangle _{3^{^{\prime }}}\left| \omega \right\rangle _{4^{^{\prime
}}}+\left| \omega +\delta \right\rangle _{1^{^{\prime }}}\left| \omega
\right\rangle _{2^{^{\prime }}}\left| \omega +\delta \right\rangle _4]\left|
\omega \right\rangle _T+  \nonumber \\
&&\lbrack \ \left| \omega \right\rangle _1\left| \omega +\delta
\right\rangle _{3^{^{\prime }}}\left| \omega \right\rangle _{4^{^{\prime
}}}+\left| \omega +\delta \right\rangle _{1^{^{\prime }}}\left| \omega
\right\rangle _{2^{^{\prime }}}\left| \omega +\delta \right\rangle _4]\left|
\omega +\delta \right\rangle _{T^{^{\prime }}}\}.  \eqnum{16}
\end{eqnarray}
Obviously, if  only one photon is detected by any one of detectors, the
state of the Eq. (16) collapses into a frequency superposition state.

\begin{equation}
\left| \omega \right\rangle _1\left| \omega +\delta \right\rangle
_{3^{^{\prime }}}\left| \omega \right\rangle _{4^{^{\prime }}}+\left| \omega
+\delta \right\rangle _{1^{^{\prime }}}\left| \omega \right\rangle
_{2^{^{\prime }}}\left| \omega +\delta \right\rangle _4.  \eqnum{17}
\end{equation}

To form a frequency-entangled GHZ state from the superposition state of Eq.
(17), one must erase all ways by which one might in principle identify true
pairs. The pair produced from one source will in general carry correlation
in polarization, energy and time. Any of these may be exploited to identify
the true sibling and hence prevent a GHZ state from forming. However,
polarization correlation can never be exploited if all photons from the two
sources carry the same polarization. This is very easy to realize. For
example, we let both sources be SPDC with type I phase matching. By AOM, the
energy correlation of true pairs(emitted by the same source) are
indistinguishable from mixed pairs (one photon from each source). The
temporal correlation can never be exploited if all four photons are produced
or detected at the same time or, more generally, if the temporal correlation
of true pairs and one of mixed pairs are indistinguishable. In order for
that, it is necessary that the coherence time of the photons is
substantially longer than the duration of the pulse. We can achieve this by
placing two narrow filters F($\omega $) and F($\omega +\delta $) in front of
detector D$_T$ and D$_{T^{^{\prime }}}$ respectively, and the bandwidths $%
\sigma _1$ of F($\omega $) and $\sigma _2$ of F($\omega +\delta $) satisfy
the relation

\begin{equation}
\sigma _p\geq \sigma _1,\sigma _2  \eqnum{18}
\end{equation}
where, $\sigma _p$ is the bandwidth of the pump pulse. By these above, a
frequency entangled GHZ state is formed, which is also a beam entangled GHZ
state. The probability of obtaining the GHZ state is about 50\%.

A frequency entangled GHZ state can also be produced from two pairs of
non-maximally frequency entangled states by the same way. For example, from

\begin{equation}
\left| \Phi \right\rangle =\cos \alpha \left| \omega \right\rangle _1\left|
\omega +\delta \right\rangle _2+\sin \alpha \left| \omega +\delta
\right\rangle _{1^{^{\prime }}}\left| \omega \right\rangle _{2^{^{\prime }}}
\eqnum{19}
\end{equation}
and

\begin{equation}
\left| \Psi \right\rangle =\cos \alpha \left| \omega \right\rangle _3\left|
\omega +\delta \right\rangle _4+\sin \alpha \left| \omega +\delta
\right\rangle _{3^{^{\prime }}}\left| \omega \right\rangle _{4^{^{\prime }}}.
\eqnum{20}
\end{equation}
The probability of getting a GHZ state is about $\sin ^2\alpha \cos ^2\alpha 
$.The extension to producing the requency entangled state of a higher number
of photons from frequency entangled states of a lower number of photons is
very easy, see Ref. [12].

\section{Conclusion}

In this paper, we show some manipulations of frequency entangled state by
AOM. As two examples, we discuss how to realize frequency entanglement
swapping and how to produce a three-photon frequency entanglement GHZ state.
These schemes can be extended to realizing frequency entanglement purifying
and to producing a frequency entangled states of multiply photons using the
schemes in Refs. [12] and [19]. All schemes are very simple, and may be
implementable in practice. The weakness of our schemes is that all proposals
are probabilisticly realizable.

This subject is supported by the National Natural Science Foundation for
Youth of China.

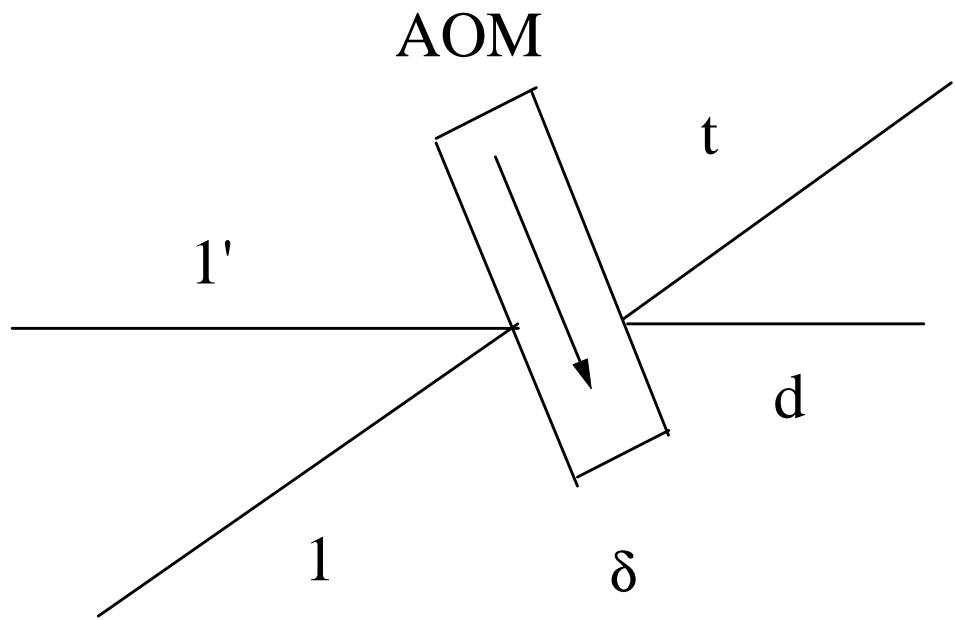
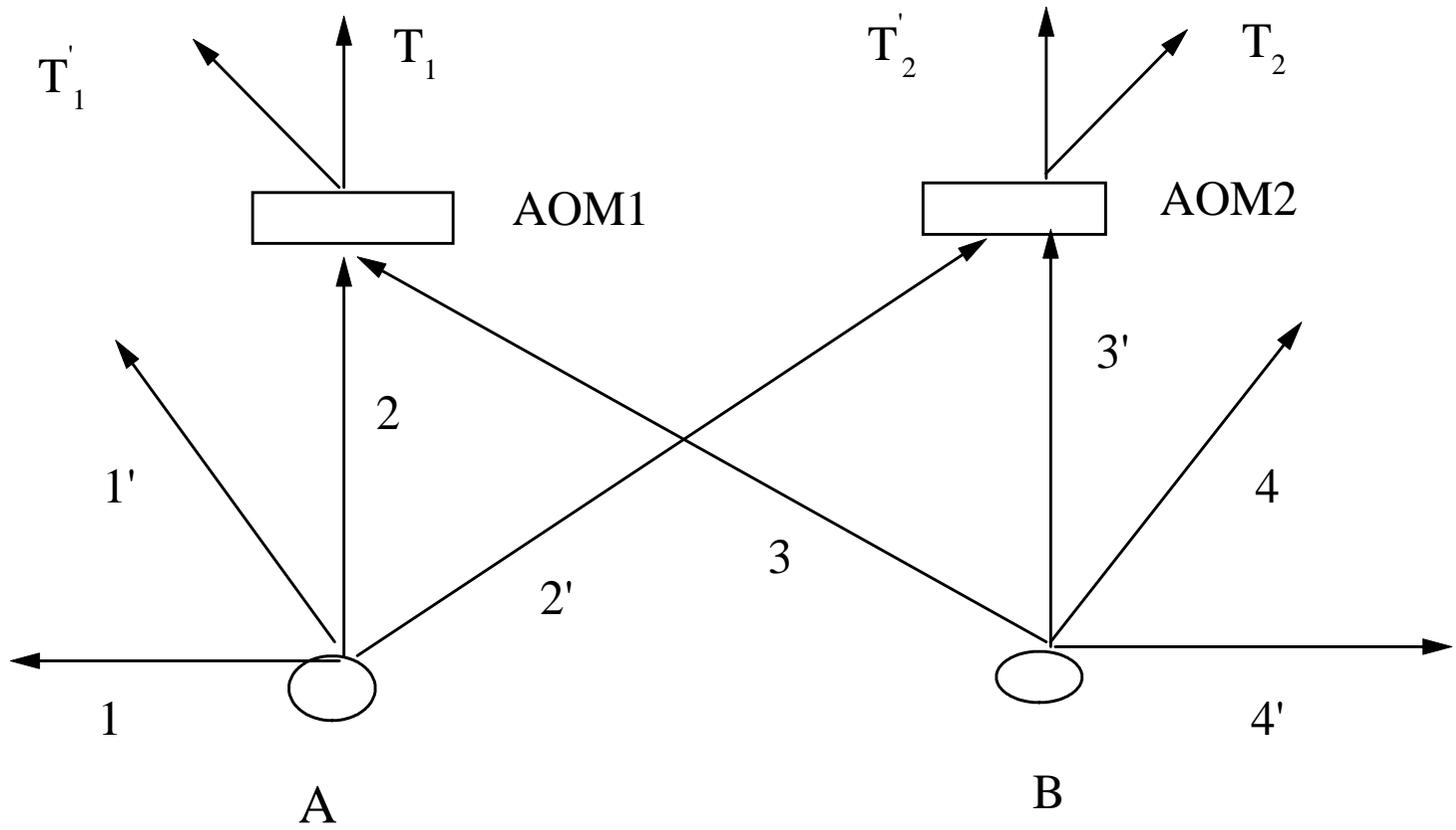
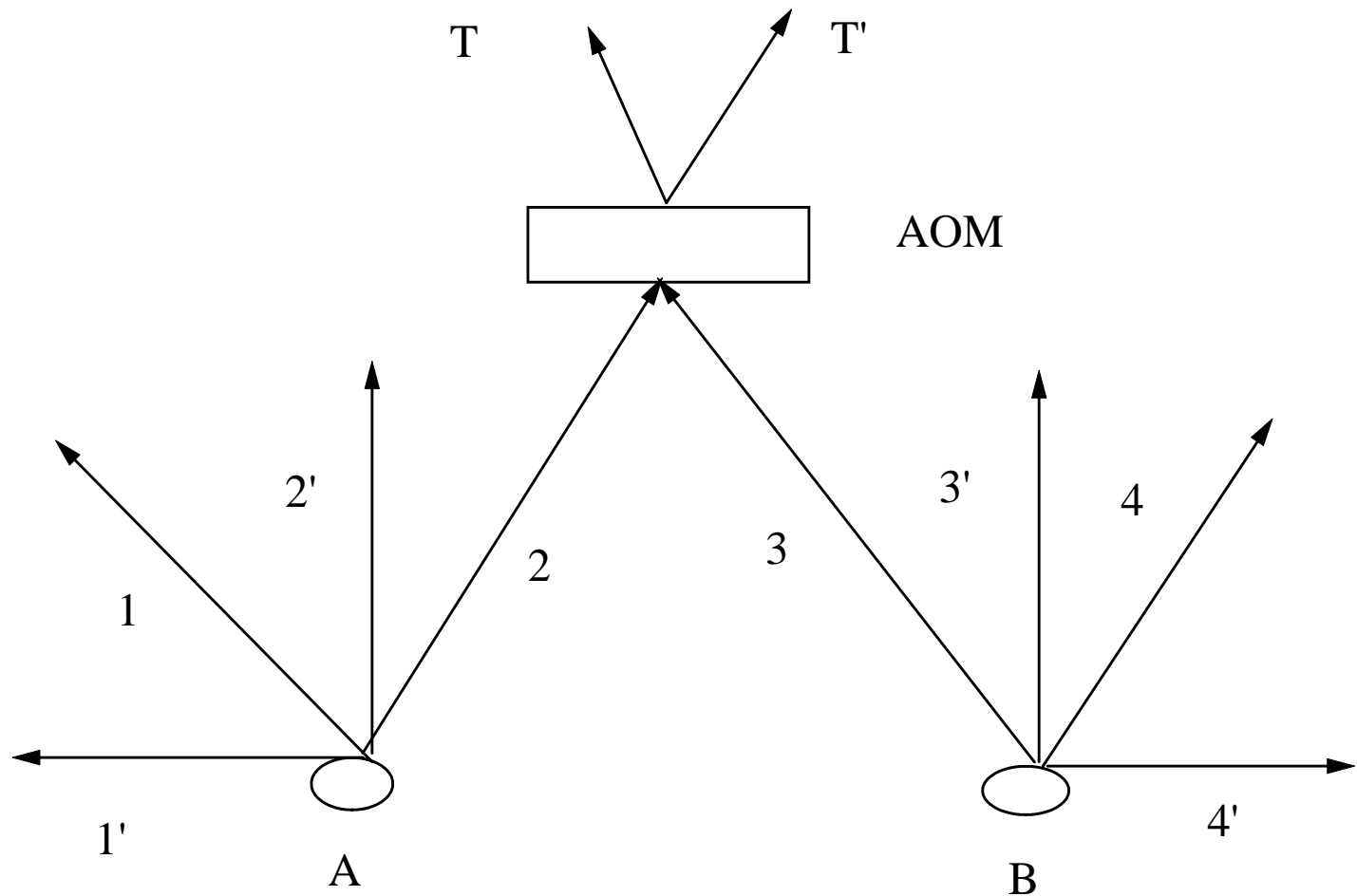
\end{document}